\documentclass[aps,showpacs,prl,twocolumn,groupedaddress]{revtex4}
\usepackage{graphicx,epsfig}
\usepackage{amsmath}
\date{\today}
\usepackage[usenames]{color}
\usepackage{ulem} 
\newcommand{\cN}{{\mathcal N}}

\begin{document}
\title{Universal heat conductance of one-dimensional channels}
%Independence of Exclusion Statistics of Heat Transport and thermodynamics
\author{Drago\c s-Victor Anghel}
\affiliation{Department of Theoretical Physics, National Institute for Physics and Nuclear Engineering--''Horia Hulubei'', 407 Atomi\c stilor street, P.O.BOX MG-6, Bucharest - Magurele, Romania}
%
%\author{}
%\affiliation{}
%
%\pacs{05.30.-d}{Quantum statistical mechanics}
%\pacs{05.30.Ch}{Quantum ensemble theory}
%\pacs{05.30.Pr}{Fractional statistics systems (anyons, etc.)}

\begin{abstract}
I analyse the transport of particles of arbitrary statistics (Bose, Fermi and fractional exclusion statistics) through one-dimensional (1D) channels. Observing that the particle, energy, entropy and heat fluxes through the 1D channel are similar to the particle, internal energy, entropy and heat capacity of a quantum gas in a two-dimensional (2D) flat box, respectively, I write analytical expressions for the fluxes at arbitrary temperatures. Using these expressions, I show that the heat and entropy fluxes are independent of statistics at any temperature, and not only in the low temperature limit, as it was previously known. From this perspective, the quanta of heat conductivity represents only the low temperature limit of the 1D channel heat conductance and is equal (up to a multiplicative constant equal to the Plank constant times the density of states at the Fermi energy) to the universal limit of the heat capacity of quantum gases. In the end I also give a microscopic proof for the universal temperature dependence of the entropy and heat fluxes through 1D channels. 
\end{abstract}

\maketitle

% \section{Introduction}\label{intro}

Rego and Kirczenow \cite{PhysRevLett.81.232.1998.Rego} and, independently, Angelescu, Cross, and Roukes \cite{SuperlatticesMicrostructures23.673.1998.Angelescu}, proved theoretically that the phonons heat conductance, $\kappa$, of a quasi one-dimensional dielectric wire in the ballistic regime in the low temperature limit, is quantized in units of 
\begin{equation}
  \kappa_0\equiv\frac{\pi^2k_B^2T}{3h}, \label{kappa_LT} %\quad \kappa = \cN_c\kappa_0, 
\end{equation}
namely $\kappa = \cN_c\kappa_0$, where $\cN_c$ is the number of phonon channels available along the wire, $h$ is the Plank constant and $k_B$ is the Boltzmann constant and $T$ is the average temperature between the ends 1 and 2 of the wire, $T=(T_1+T_2)/2$, assuming that $|T_1-T_2|/T\ll 1$. 

These results have been experimentally confirmed in Refs. \cite{Nature404.974.2000.Schwab,Nature.444.187.2006.Meschke} and have been extended by Rego and Kirczenow in Ref. \cite{PhysRevB.59.13080.1999.Rego}, where they showed that the same quantization rule applies to the heat conductance of particles of any statistics. 

Let us consider a two-terminal transport experiment in which the reservoirs 1 and 2 are connected by a quasi 1D wire. The temperatures and chemical potentials in the two reservoirs will be denoted by $T_{i}$ and $\mu_{i}$, respectively ($i=1,2$). The particles in the system may be bosons, fermions, or may obey fractional exclusion statistics (FES) of parameter $\alpha$ \cite{PhysRevLett.67.937.1991.Haldane,PhysRevLett.73.922.1994.Wu,PhysRevLett.104.198901.2010.Anghel,PhysRevLett.104.198902.2010.Wu,EPL.87.60009.2009.Anghel}. Then the Landauer formula for the particle and heat fluxes between the two reservoirs are \cite{PhysRevB.59.13080.1999.Rego,PhysRep.395.159.2004.Blencowe}
\begin{subequations}\label{IUgen}
\begin{eqnarray}
  I &=& \sum _{n=1}^{\cN_c}\int_{0}^{\infty}\frac{dk}{2\pi}v_{n}(k)\left[\eta_{1}(k)-\eta_{2}(k)\right]\zeta_n(k) 
  \label{Igen} \\
  \dot U &=& \sum _{n=1}^{\cN_c}\int_{0}^{\infty}\frac{dk}{2\pi}\epsilon v_{n}(k)\left[\eta_{1}(k)-\eta_{2}(k)\right]\zeta_n(k)
  \label{Ugen}
\end{eqnarray}
\end{subequations}
where $v_{n}(k)$ is the group velocity of the particles of momentum $k$, $\eta_{i}(k)$ denotes the thermal particle population in the reservoir $i$ and, finally, $\zeta_n(k)$ is the particle transmission coefficient through the wire. The summation is taken over the 1D (available) channels along the wire.

Since the particle group velocity is $v_{n}(k)=\hbar^{-1}(d\epsilon(k)/dk)$, and assuming further that $\zeta_n(k)\equiv\zeta_n$ is independent of $k$, Eqs. (\ref{IUgen}) get the simple form 
\begin{subequations}\label{IUgen1}
\begin{eqnarray}
  I &=& \frac{\zeta_n}{h}\sum _{n=1}^{\cN_c}\int_{\epsilon_n(0)}^{\infty}d\epsilon\left[\eta_{1}(\epsilon)-\eta_{2}(\epsilon)\right]
  \label{Igen1} \\
  \dot U &=& \frac{\zeta_n}{h}\sum _{n=1}^{\cN_c}\int_{\epsilon_n(0)}^{\infty}d\epsilon\,\epsilon\left[\eta_{1}(\epsilon)-\eta_{2}(\epsilon)\right]
  \label{Ugen1}
\end{eqnarray}
\end{subequations}
where $\epsilon_n(0)$ is the lowest energy level in the channel $n$. 

To calculate the heat conductivity, $\kappa\equiv\dot U/(T_1-T_2)$, one introduces in (\ref{IUgen1}) the FES populations \cite{PhysRevB.59.13080.1999.Rego,PhysRevLett.73.922.1994.Wu,EPL.90.10006.2010.Anghel}, 
\begin{subequations}\label{EqFES}
\begin{eqnarray}
  &&\eta_{i}(\epsilon)\equiv\eta_\alpha(\epsilon,\mu,T) = \left[w_\alpha(\mu,T)+\alpha\right], \label{Eq_eta} \\
% \end{equation}
%
%with $w$ given by the equation
%
% \begin{equation}
  &&w_\alpha^{\alpha}(\mu,T)[1+w_\alpha(\mu,T)]^{1-\alpha}=\exp[\beta(\epsilon-\mu)], \label{Eq_w}
\end{eqnarray}
\end{subequations}
where $\beta\equiv 1/(k_BT)$ ($\alpha=0$ and 1 correspond to bosons and fermions, respectively). By doing so, Rego and Kirczenow calculated the low temperature limit of $\kappa$ and observed that it is independent of $\alpha$ and therefore the quantization relation (\ref{kappa_LT}) holds for particles of any statistics \cite{PhysRevB.59.13080.1999.Rego,PhysRep.395.159.2004.Blencowe}. Nevertheless, a physical understanding of this mathematical result is still missing \cite{PhysRep.395.159.2004.Blencowe}. 

In this letter I extend the results of Rego and Kirczenow by showing that the heat conductivity of a 1D channel is independent of the statistics of particles at any temperature and I will provide a microscopic explanation for this result. This is done by observing that there is a close similarity between the stationary heat and particle transport in 1D and equilibrium thermodynamics in 2D (or, in systems with constant single-particle density of states, DOS \cite{ProcCambrPhilos42.272.1946.Auluc,PhysRev.135.A1515.1964.May,PhysRevE.55.1518.1997.Lee,PhysA304.421.2002.Lee,JPA35.7255.2002.Anghel,RJP.54.281.2009.Anghel,JPA38.9405.2005.Anghel}). The quanta of heat conductance is then nothing but the low temperature limit of the 1D heat conductance. Moreover, I calculate the entropy current through the channel and I show that this is also independent of statistics at any temperature as long as the particle current is zero.

% \section{Universality of the quantum of heat conductance}

Let us focus on systems with only one channel, $n$. We split the particle and heat fluxes into two parts, one coming from the reservoir 1 and one from the reservoir 2, $I_n\equiv I_{n,1}(\mu_1,T_1)-I_{n,1}(\mu_2,T_2)$ and $\dot U_n\equiv\dot U_{n,1}(\mu_1,T_1)-\dot U_{n,1}(\mu_2,T_2)$, where 
\begin{subequations}\label{IUnsplit}
\begin{eqnarray}
  I_{n,1}(\mu_n,T_n) &\equiv& \frac{\zeta_n}{h}\int_{\epsilon_n(0)}^{\infty}\eta_{\alpha}(\epsilon,\mu_n,T_n)d\epsilon \label{Insplit} \\
  \dot U_{n,1}(\mu_n,T_n) &\equiv& \frac{\zeta_n}{h}\int_{\epsilon_n(0)}^{\infty}\epsilon\eta_{\alpha}(\epsilon,\mu_n,T_n)d\epsilon 
  \label{Unsplit}
\end{eqnarray}
\end{subequations}
We observe that $I_{n,1}(\mu_n,T_n)$ and $\dot U_{n,1}(\mu_n,T_n)$ have exactly the same expressions as the particle number, $N_\alpha(\mu_n,T_n)$, and internal energy, $U_\alpha(\mu_n,T_n)$, respectively, of a FES system of parameter $\alpha$ in equilibrium at temperature $T$ and chemical potential $\mu$, which has a constant DOS, $\sigma=\zeta_n/h$. Therefore we can apply here directly the results we obtained for these latter systems in Ref. \cite{JPA35.7255.2002.Anghel}. To specify the notations, let us briefly review the results of Ref. \cite{JPA35.7255.2002.Anghel} which are of interest here. 

If a FES system of parameter $\alpha$ and density of states $\sigma=\zeta_n/h$, in equilibrium at temperature $T$, contains $N_\alpha$ particles, then we define the statistics independent parameter $y_0$ by the relation 
\begin{subequations}\label{def_N_y0_mu}
\begin{equation}
  N_\alpha \equiv k_BT\sigma\log(1+y_0). \label{N}
\end{equation}
%
% We observe that $y_0$ is a monotonically decreasing function of $T$, with the property that $0=y_0(N_\alpha,T=\infty)\le y_0(N,T)\le y_0(N_\alpha,T=0)=\infty$. 
%
Knowing $y_0$, one may calculate the chemical potential, $\mu$, from the equation 
\begin{equation}
  (1+y_0)^{1-\alpha}/y_0 = e^{-\beta\mu}  \label{defy0}
\end{equation}
and from Eqs. (\ref{N}) and (\ref{defy0}) we observe that 
\begin{equation}\label{mu}
\exp{[(\mu - \alpha N_\alpha/\sigma)/k_{\rm B} T]} = 1- \exp{[-N_\alpha/(\sigma k_{\rm B}T)]}. 
\end{equation}
\end{subequations}
If we identify the (generalized) Fermi energy as $\epsilon_{\rm F}\equiv\lim_{T\to 0}\mu = \alpha N_\alpha/\sigma$ we observe that $\mu-\epsilon_{\rm F}$ is also independent of $\alpha$, or, \textit{vice-versa}, $\mu-\epsilon_{\rm F}$ determine uniquely $N_\alpha$, for any $\alpha$.

In these notations, the grandcanonical potential and internal energy, $\Omega_\alpha$ and $U_\alpha$ are \cite{JPA35.7255.2002.Anghel}
\begin{eqnarray}\label{echiv} 
\Omega_\alpha = -U_\alpha &=& (k_{\rm B}T)^2 \sigma\left[\frac{1-\alpha}{2} 
\log^2{(1+y_0)} + Li_2(-y_0)\right] \nonumber \\
&& = \frac{1-\alpha}{2}\frac{N_\alpha^2}{\sigma} + (k_{\rm B} T)^2 \sigma Li_2(-y_0) ,
\end{eqnarray}
where $Li_2$ is the Euler's dilogarithm, $Li_2(z) = \sum_{i=1}^{\infty}z^k/k^2$ \cite{Lewin:book}. 
%
% \begin{equation}
%   Li_2(z) = \sum_{i=1}^{\infty}\frac{z^k}{k^2}. \label{defLi2}
% \end{equation}
%
The fact that, at constant $N_\alpha$, the temperature dependent part of $\Omega_\alpha$ and $U_\alpha$, i.e. $(k_{\rm B} T)^2\sigma Li_2(-y_0)$, is independent of $\alpha$ is an expression of the thermodynamic equivalence of quantum gases of the same, constant DOS \cite{ProcCambrPhilos42.272.1946.Auluc,PhysRev.135.A1515.1964.May,PhysRevE.55.1518.1997.Lee,PhysA304.421.2002.Lee,JPA35.7255.2002.Anghel}. 

From Eqs. (\ref{echiv}) and (\ref{N}), one can obtain the entropy and the heat capacity \cite{JPA35.7255.2002.Anghel},
\begin{eqnarray}
S %&\equiv& \left.\frac{d\Omega_\alpha}{dT}\right|_{\mu\ {\rm constant}} \nonumber \\
&=& -k_{\rm B}^2 T \sigma [2Li_2(-y_0) + \log{(1+y_0)}\log{y_0}] 
\label{entropia} \\
%&=& -k_{\rm B}^2 T C [Li_2(-y_0) - Li_2(1+y_0) - i\pi \log{(1+y_0)} + 
%\frac{\pi^2}{6}] \nonumber \\
C_{\rm V} %&\equiv& \left.\frac{dU_\alpha}{dT}\right|_{N_\alpha\ {\rm constant}} \nonumber \\
&=& - \frac{N_\alpha^2}{T\sigma}\frac{1+y_0}{y_0}- 2k_{\rm B}^2 
T\sigma Li_2(-y_0) , \label{cv} 
\end{eqnarray}
which are both independent of $\alpha$. Since, according to Eq. (\ref{N}), $\lim_{T\to0}y_0=\infty$, using the asymptotic behavior of the dilogarithm, $Li_2(-y_0) \sim -[\pi^2/6 + \log^2(y_0)/2]$, one can recover in the low temperature limit the universal asymptotic expression for the heat capacity of the system, namely \cite{Stone:book} 
\begin{equation}
  C_V\sim(\pi^2/3)k_B^2T\sigma. \label{cv_limT0}
\end{equation}

Now we have all the ingredients and we can transcribe the formalism above into a formalism for the particle and heat transport along the wire. For this, we identify $I_{n,1}(\mu_n,T_n)$ with $N_\alpha(\mu_n,T_n)$, and we introduce $y_{i0}$ by
\begin{subequations}\label{IUpoly}
\begin{eqnarray}
  &&I_{n,1}(\mu_n,T_n) \equiv \frac{\zeta_nk_BT_n}{h}\log(1+y_{i0}), \label{In} \\
% \end{equation}
%
% and
%
% \begin{equation}
  && (1+y_{i0})^{1-\alpha}/y_{i0} = e^{-\beta_n\mu_n},  \label{yin} 
\end{eqnarray}
%\end{subequations}
%
% From the last two equations we write an expression similar to Eq. (\ref{mu}),
%
\begin{equation}
  \exp{[(\mu - \alpha h I_{n,1}/\zeta_n)/k_{B} T_n]} = 1- \exp{[-hI_{n,1}/(\zeta_n k_{B}T_n)]},  \label{muI}
\end{equation}
in analogy to Eqs. (\ref{def_N_y0_mu})

From the equations above and observing that $\dot U_{n,1}(\mu_n,T_n)$ has an expression similar to that of $U_\alpha(\mu_n,T_n)$, we obtain 
\begin{eqnarray}
\dot U_{n,1} %&=& (k_{\rm B}T)^2 \frac{\zeta_n}{h}\left[\frac{1-\alpha}{2}\log^2{(1+y_0)} + Li_2(-y_0)\right] \nonumber \\ %\label{dotUechiv1}
  &=& -\frac{1-\alpha}{2}\frac{hI_{n,i}^2}{\zeta_n} - (k_{\rm B} T_n)^2 \frac{\zeta_n}{h} Li_2(-y_{i0}) .\label{dotUechiv2} 
\end{eqnarray}
\end{subequations}
From Eqs. (\ref{IUpoly}) we calculate
%
% \begin{subequations}\label{dIU} 
% \begin{eqnarray}
%   dI_{n,1} &=& \frac{\partial I_{n,1}}{\partial T}dT + \frac{\partial I_{n,1}}{\partial \mu}d\mu, \label{dI} \\
%   d\dot U_{n,1} &=& \frac{\partial \dot U_{n,1}}{\partial T}dT + \frac{\partial \dot U_{n,1}}{\partial \mu}d\mu. \label{dU} 
% \end{eqnarray}
% \end{subequations}
%
% 
% From Eq. (\ref{muI}) we obtain after some algebra 
%
\begin{subequations} \label{dIdTdmu}
\begin{eqnarray}
  \frac{\partial I_{n,1}}{\partial T} &=& \frac{I_{n,1}}{T}-\frac{\zeta_n\mu}{hT}\frac{1-\exp\left[-\frac{hI_{n,1}}{\zeta_n k_BT}\right]}{\alpha+(1-\alpha)\exp\left[-\frac{hI_{n,1}}{\zeta_n k_BT}\right]} \nonumber \\
  &=& \frac{\zeta_n k_B}{h}\log(1+y_{i0}) - \frac{\zeta_n\mu}{hT}\frac{y_{i0}}{1+\alpha y_{i0}}, \label{dIdT}
\end{eqnarray}
\begin{equation}
  \frac{\partial I_{n,1}}{\partial \mu} = \frac{\frac{\zeta_n}{h}\left[1-\exp\left(-\frac{hI_{n,1}}{\zeta_n k_BT}\right)\right]}{\alpha+(1-\alpha)\exp\left(-\frac{hI_{n,1}}{\zeta_n k_BT}\right)}
  = \frac{\zeta_n y_{i0}}{h(1+\alpha y_{i0})}.
  \label{dIdmu}
\end{equation}
\end{subequations}
%
% 
% To calculate $\partial \dot U_{n,1}/\partial T$ and $\partial \dot U_{n,1}/\partial\mu$ we make use of the Eqs. (\ref{IUpoly}) and (\ref{dIdTdmu}) to write
% %
% \begin{equation} \label{dy0dmudT}
%   \left.\frac{d y_0}{dT}\right|_{\mu} = \frac{\mu y_0(1+y_0)}{k_BT^2(1+\alpha y_0)} 
%   \ {\rm and}\ 
%   \left.\frac{d y_0}{d\mu}\right|_{T} = \frac{y_0(1+y_0)}{k_BT(1+\alpha y_0)}
% \end{equation}
% %
% and using these, we obtain
%
\begin{subequations} \label{ddotUdTdmu}
\begin{eqnarray}
\frac{\partial \dot U_{n,1}}{\partial T} &=& -(1-\alpha)\frac{hI_{n,1}}{\zeta_n}\frac{\partial I_{n,1}}{\partial T} - 2k_{\rm B}^2 T_n \frac{\zeta_n}{h} Li_2(-y_{i0}) \nonumber \\
&& + \frac{\mu I_{n,1}(1+y_{i0})}{T(1+\alpha y_{i0})}, \label{ddotUdT}
\\ 
\frac{\partial \dot U_{n,1}}{\partial\mu} &=&-(1-\alpha)\frac{hI_{n,1}}{\zeta_n}\frac{\partial I_{n,1}}{\partial\mu} - 2k_{\rm B}^2 T_n \frac{\zeta_n}{h} Li_2(-y_{i0}) \nonumber \\
&& + \frac{I_{n,1}(1+y_{i0})}{1+\alpha y_{i0}}, \label{ddotUdmu}
\end{eqnarray}
\end{subequations}
%
% where I used the identity $-dLi_2(-y_{i0})/dy_{i0}=\log(1+y_{i0})/y_{i0}$ obtained from Eq. (\ref{defLi2}). 

To calculate the low temperature approximations, we use the fact that $\lim_{T\to0}y_{i0}=\infty$ for any $\alpha$ and obtain
\begin{subequations}\label{lowTexpr1}
\begin{eqnarray}
  y_{i0,\alpha>0} &\stackrel{T\to0}{\sim}& e^{\frac{\mu}{\alpha k_BT}}+\frac{1-\alpha}{\alpha},\label{lowTy0} \\
  I_{n,1,\alpha>0} &\stackrel{T\to0}{\sim}& \frac{\zeta_n\mu}{h\alpha} + \frac{(1-\alpha)\zeta_n k_BT}{\alpha h}e^{-\frac{\mu}{\alpha k_BT}} \label{lowTI} \\
  \dot U_{n,1,\alpha>0} &\stackrel{T\to0}{\sim}& \frac{\zeta_n\mu^2}{2h\alpha} +\frac{\pi^2\zeta_n(k_BT)^2}{6h} +\frac{(1-\alpha)\zeta_nk_BT\mu}{\alpha h} 
  \nonumber \\
  && \times e^{-\frac{\mu}{\alpha k_BT}} \label{lowTdotU}
\end{eqnarray}
for $\alpha>0$, whereas for $\alpha=0$ we have 
\begin{eqnarray}
  y_{i0,\alpha=0} &\stackrel{T\to0}{\sim}& -\frac{k_BT}{\mu}-\frac{1}{2},\label{lowTy0a0} \\
  I_{n,1,\alpha=0} &\stackrel{T\to0}{\sim}& -\frac{\zeta_nk_BT}{h}\left[\log\left(-\frac{\mu}{k_BT}\right)+\frac{\mu}{2k_BT}\right] \label{lowTIa0} \\
  \dot U_{n,1,\alpha=0} &\stackrel{T\to0}{\sim}& \frac{\pi^2\zeta_n(k_BT)^2}{6h} -\frac{\zeta_nk_BT\mu}{h}\log\left[-\frac{\mu}{k_BT}\right] \label{lowTdotUa0}
\end{eqnarray}
\end{subequations}

From Eqs. (\ref{lowTexpr1}) we calculate the derivatives of $I_{n,1}$ and $\dot U_{n,1}$ at low temperatures:
\begin{subequations}\label{lowTderivs}
  \begin{eqnarray}
    \frac{\partial I_{n,1,\alpha>0}}{\partial T} &\stackrel{T\to0}{\sim}& \frac{(1-\alpha)\zeta_nk_B}{\alpha h}\left[\frac{\mu}{\alpha k_BT}+1\right]e^{-\frac{\mu}{\alpha k_BT}} \label{lowTdIpdtT0} \\
    \frac{\partial I_{n,1,\alpha>0}}{\partial\mu} &\stackrel{T\to0}{\sim}& \frac{\zeta_n}{h\alpha} - \frac{(1-\alpha)\zeta_n}{\alpha^{2} h}e^{-\frac{\mu}{\alpha k_BT}}  \label{lowTdIpdmuT0} \\
    \frac{\partial \dot U_{n,1,\alpha>0}}{\partial T} &\stackrel{T\to0}{\sim}& \frac{\pi^2\zeta_nk_B^2T}{3h} +\frac{(1-\alpha)\zeta_nk_BT\mu}{\alpha h} \nonumber \\
    && + \frac{(1-\alpha)\zeta_nk_B\mu}{\alpha h}\left[\frac{\mu}{\alpha k_BT}+1\right]e^{-\frac{\mu}{\alpha k_BT}} \label{lowTddotUpdtT0} \\
    \frac{\partial \dot U_{n,1,\alpha>0}}{\partial \mu} &\stackrel{T\to0}{\sim}& \frac{\zeta_n\mu}{h\alpha} -\frac{(1-\alpha)\zeta_nk_BT}{\alpha h}\left[\frac{\mu}{\alpha k_BT}-1\right] \nonumber \\ 
    && \times e^{-\frac{\mu}{\alpha k_BT}} \label{lowTddotUpdmuT0}\\
    %%%%%%%%%%%%%%%%%%%%%%%%%%%%%%%%%%%%%%%%%%%%%%%%%%%%%%%%%%%%%%%%%
    \frac{\partial I_{n,1,\alpha=0}}{\partial T} &\stackrel{T\to0}{\sim}& \frac{\zeta_nk_B}{\alpha h}\left[-\log\left(-\frac{\mu}{\alpha k_BT}\right)+1\right] \label{lowTdIpdta0} \\
    \frac{\partial I_{n,1,\alpha=0}}{\partial\mu} &\stackrel{T\to0}{\sim}& -\frac{\zeta_n k_BT}{h\mu}\left(1+\frac{\mu}{2 k_BT}\right) \label{lowTdIpdmuT0a0} \\
    \frac{\partial \dot U_{n,1,\alpha=0}}{\partial T} &\stackrel{T\to0}{\sim}& \frac{\pi^2\zeta_nk_B^2T}{3h} +\frac{\zeta_nk_B^2T}{h}\frac{\mu}{k_BT} \nonumber \\
    && \times\left[-\log\left(-\frac{\mu}{k_BT}\right)-1\right] \label{lowTddotUpdta0} \\
    \frac{\partial \dot U_{n,1,\alpha=0}}{\partial \mu} &\stackrel{T\to0}{\sim}& \frac{\zeta_nk_BT}{h}\left[-\log\left(-\frac{\mu}{k_BT}\right)-1\right] \label{lowTddotUpdmua0}
  \end{eqnarray}
\end{subequations}
We observe that the results (\ref{lowTderivs}) for $\alpha>0$ coincide  in the lowest order approximation with the results (16)-(19) of Ref. \cite{PhysRevB.59.13080.1999.Rego}, but they are different in general for $\alpha=0$. For example $\lim_{T\to0}\partial I_{n,1}/\partial T=\infty$ for $\alpha=0$, whereas for $\alpha>0$ $\lim_{T\to0}\partial I_{n,1}/\partial T=0$. The only result that remains the same in the lowest order approximation for both, $\alpha>0$ and $\alpha=0$, is $\partial\dot U_{n,1}/\partial T=\zeta_n\pi^2k_B^2T/(3h)$, which is the quanta of heat conductance if we set $\zeta_n=1$.

Nevertheless, in real heat conductance measurements the heat transport takes place without particle transport. Therefore the derivative $(\partial\dot U_{n,1}/\partial T)_\mu$ (taken at constant $\mu$) is not the quantity of interest for us. The quantity which represents the heat conductivity is actually $(d\dot{U}_{n,1}/dT)_{I_{n}=0}\equiv(d\dot{U}_{n,1}/dT)_{I_{n,1}}$, which is the derivative of $\dot U_{n,1}$ with respect to $T$, at constant $I_{n,1}(\mu,T)$. In these conditions the variations of $T$ and $\mu$ between the two reservoirs are related by
\begin{equation}
  \left.\frac{d\mu}{dT}\right|_{I_{n,1}} = -\left.\frac{\partial I_{n,1}}{\partial T}\right|_{\mu}\left(\left.\frac{\partial I_{n,1}}{\partial \mu}\right|_{T}\right)^{-1}. \label{dmudT_I}
\end{equation}
Plugging Eq. (\ref{dmudT_I}) into the expression for $(d\dot U_{n,1}/dT)_{I_{n,1}}$, we obtain 
\begin{equation}
  \kappa=\left.\frac{d\dot U_{n,1}}{dT}\right|_{I_{n,1}} = \frac{\partial\dot U_{n,1}}{\partial T}-\frac{\partial\dot U_{n,1}}{\partial \mu}\left.\frac{\partial I_{n,1}}{\partial T}\right|_{\mu}\left(\left.\frac{\partial I_{n,1}}{\partial \mu}\right|_{T}\right)^{-1} \label{kappa_gen}
\end{equation}
which, if we replace $I_{n,1}$ by $N_\alpha$ and $\dot U_{n,1}$ by $U_\alpha$, becomes identical to the expression for the heat capacity, $C_V$ (\ref{cv}). Therefore we can transcribe Eq. (\ref{cv}) for the heat conductivity as 
\begin{equation}
  \kappa_{n} = - \frac{I_{n,1}^2}{T\sigma}\frac{1+y_{i,0}}{y_{i,0}}-2k_{\rm B}^2T\sigma Li_2(-y_{i,0}) , \label{kappa_univ} 
\end{equation}
which is independent of $\alpha$ at any temperature. In the low temperature limit Eq. (\ref{kappa_univ}) becomes the universal asymptotic expression, $\kappa_0$ (\ref{kappa_LT}), which is equivalent to (\ref{cv_limT0}) from equilibrium thermodynamics. 

The result (\ref{kappa_LT}) was obtained before \cite{PhysRevLett.81.232.1998.Rego,SuperlatticesMicrostructures23.673.1998.Angelescu,PhysRevB.59.13080.1999.Rego,PhysRep.395.159.2004.Blencowe} without imposing the condition $dI_{n,1}=0$ due to the fact that the second term at the right hand side of Eq. (\ref{kappa_gen}) converges to zero at $T\to0$, as one can readily check from the asymptotic expressions (\ref{lowTderivs}).

In Ref. \cite{PhysRep.395.159.2004.Blencowe} Blencowe analyzed also the entropy flux through the 1D channel in the low temperature limit and observed that it is independent of $\alpha$. Let's analyze it here from the perspective of Ref. \cite{JPA35.7255.2002.Anghel} and prove that it is independent of $\alpha$ at any temperature. 

The entropy flux from one of the reservoirs may be calculated in a way similar to the calculation of heat and particle fluxes:
\begin{eqnarray}
  \dot S &=& \frac{k_B\zeta_n}{h}\int_{0}^{\infty}\big\{[1+(1-\alpha)n(\epsilon)]\log[1+(1-\alpha)n(\epsilon)]\nonumber \\ 
  && -n(\epsilon)\log[n(\epsilon)]-[1-\alpha n(\epsilon)]\log[1-\alpha n(\epsilon)]\big\}d\epsilon \nonumber 
  \\
  &=& -\frac{k_{\rm B}^2 T\zeta_n}{h} [2Li_2(-y_{i,0}) + \log{(1+y_{i,0})}\log{y_{i,0}}] \label{entropia_flux} ,
\end{eqnarray} 
and represents the flux of the number of configurations of particle populations, $\{n(\epsilon)\}$, from one reservoir to the other. But Eq. (\ref{entropia_flux}) is identical in form with Eq. (\ref{entropia}) and therefore it is independent of $\alpha$ at any temperature.

To clarify the microscopic reason for which the 1D entropy and heat fluxes are independent of statistics, we use Eqs. (\ref{IUpoly}) and the relation $Li_2(x)+Li_2[-x/(1-x)]=-\frac{1}{2}\log^2(1-x)$ \cite{Lewin:book} to write 
\begin{eqnarray}
  \dot U_{n,1}(\mu_i,T_i) &=& \frac{\alpha hI_{n,1}^2}{2\zeta_n} + \frac{\zeta_n(k_BT_i)^2}{h}Li_2\left(\frac{y_{0,i}}{y_{0,i}+1}\right) \nonumber \\
  &\equiv& \dot U_{0,n,1}(I_{n,1}) + \dot U_{B,n,1}(I_{n,1},T_i)  \label{def_dotUB}
\end{eqnarray}
where $\dot U_{0,n,1}$ is the energy flux at zero temperature and $\dot U_{B,n,1}$ is the excitation energy flux--in a Bose gas, $\dot U_{0,n,1}\equiv0$ and $\dot U_{n,1}\equiv\dot U_{B,n,1}$, hence the notation. As one can see directly from Eq. (\ref{def_dotUB}), $\dot U_{B,n,1}$ is independent of $\alpha$ at any temperature and therefore at zero net current, $I_{n,1}(\mu_1,T_1)-I_{n,1}(\mu_2,T_2)=0$, the heat flux is equal to the difference between the excitation energy fluxes, 
\begin{equation}
  \dot U_n = \dot U_{B,n,1}(I_{n,1},T_2)-\dot U_{B,n,1}(I_{n,1},T_1), \label{dotUN_B}
\end{equation}
which is independent of $\alpha$ at any $T$. 

The statistics independence of $\dot U_{B,n,1}(I,T)$ at any $T$ may be interpreted microscopically also starting from the analogy with the equilibrium thermodynamics of systems of constant DOS. In Ref. \cite{JPA35.7255.2002.Anghel} it was proven that for systems of particles of different statistics, but the same, constant DOS and particle number, there is a one-to-one mapping between configurations of particle populations, $\{n(\epsilon)\}$, which have the same excitation energy. This implies that the canonical partition functions of such systems are independent of $\alpha$ and therefore all their canonical thermodynamics is independent of statistics, including the entropy and the heat capacity. 

\begin{figure}[t]
  %\fbox{
  \includegraphics[width=6cm]{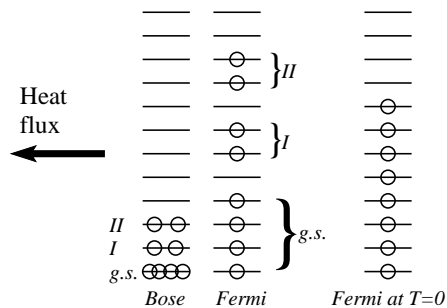}
  %}
\caption{The microscopic analysis of the statistics independence of the heat conductivity: there is a one-to-one correspondence between micro-configurations of particles of different statistics which have the same excitation energy, $\dot U_{B,n,1}$, and therefore carry the same heat-fluxes.}
\label{B_F_flux}
\end{figure} 

The same argument can be transcribed for energy fluxes in 1D channels. If we have two gases, one of parameter $\alpha$ and another of parameter $\alpha'$, both gases carrying the same particle flux, $I_{n,1}$, through a 1D channel, then one can establish a one-to-one correspondence between configurations of particle populations in the two gases, with the same $\dot U_{B,n,1}$. Two such configurations, one of bosons and one of fermions, are shown in Fig. \ref{B_F_flux}. This implies that both, the excitation energy flux and the entropy flux (which is determined by the flux of the number of configurations), are independent of statistics. 

%\section{Conclusions}

In conclusion I showed that the particle, energy, entropy and heat fluxes through a 1D channel are analogue to the particle number, internal energy, entropy and the heat capacity of a gas of constant density of states. Using this analogy, I wrote analytical expressions for all the fluxes and their derivatives with respect to the chemical potential and temperature, I calculated their asymptotic expressions in the limit $T\to0$, and I showed that the heat and entropy fluxes are independent of the statistics of the particles involved in the transport at any temperature, not only when $T\to0$, as it was known before. Using a construction I introduced in Ref. \cite{JPA35.7255.2002.Anghel}, I showed what is the microscopic reason for the independence of statistics of the constituent particles for heat and entropy fluxes in 1D channels.

%\section*{Acknowledgements}

The financial support from the Romanian National Authority for Scientific Research grant PN 09370102, and the Romanian, IFIN-HH–JINR collaboration grants 4027-3-10/11 and N4006 is gratefully acknowledged.

%\bibliography{/home/dragos/general}
%\bibliographystyle{unsrt}

\end{document}